\documentclass[aps,prep,preprintnumbers,amsmath,amssymb,nofootinbib]{revtex4}
\usepackage{amssymb}
\usepackage{lscape}
\usepackage{amsmath}
\usepackage{bm}
\usepackage{graphicx}
\usepackage{epsfig}
\begin{document}
\title{Towards relativistic quantum geometry}
\author{$^{2}$ Luis Santiago Ridao,  $^{1,2}$ Mauricio Bellini
\footnote{E-mail address: mbellini@mdp.edu.ar} }
\address{$^1$ Departamento de F\'isica, Facultad de Ciencias Exactas y
Naturales, Universidad Nacional de Mar del Plata, Funes 3350, C.P.
7600, Mar del Plata, Argentina.\\
$^2$ Instituto de Investigaciones F\'{\i}sicas de Mar del Plata (IFIMAR), \\
Consejo Nacional de Investigaciones Cient\'ificas y T\'ecnicas
(CONICET), Mar del Plata, Argentina.}

\begin{abstract}
We obtain a gauge-invariant relativistic quantum geometry by
using a Weylian-like manifold with a geometric scalar
field which provides a gauge-invariant relativistic quantum theory in which the algebra of the Weylian-like field depends on observers.
An example for a Reissner-Nordstr\"om black-hole is studied.
\end{abstract}
\maketitle

\section{Introduction}

The study of geometrodynamics was introduced by Wheeler in the 50's decade in order to describe particle as
geometrical topological defects in a relativistic framework\cite{Wh}, and, in the last years has becoming a very intensive subject of research\cite{sec}.
In the last decades Loop Quantum Gravity (LQG) have provided a picture of the quantum geometry of space, thanks in part to the theory of spin networks\cite{aa}. The concept of spin foam is intended to serve as a similar picture for the quantum geometry of spacetime. LQG is a theory that attempts to describe the quantum properties of the universe and gravity. In LQG the space can be viewed as an extremely fine favric of finite loops. These networks of loops are called spin networks. The evolution of a spin network over time is called a spin foam. The more traditional approach to LQG is the canonical LQG, and there is a newer approach called covariant LQG, more commonly called spin foam theory. However, at the present time, it is not possible to realize a consistent quantum gravity theory which leads to the unification of gravitation with the other forces.
One of the problems relies in the impossibility of constructing a non-perturbative gauge-invariant formalism which can describe intense quantised gravitational fields. This will be the subject
of this letter, but using a new kind of connections that do not preserve the norm of the vectors. Therefore, we shall call these manifolds as Weylian-like. Other remarkable characteristic of these manifolds is, as in Weylian manifolds, that the variation of the tensor metric is nonzero. To do it we shall define a geometrical scalar field $\sigma$ that drives a geometrical displacement from a Riemannian manifold (on which we define the background), to a Weylian-like manifold where we represent the dynamics of the quantum geometry.

In the following issues
of this section we shall revise the procedure for the minimization of the EH action, without making emphasis on the structure of the manifolds, but in the gauge transformations of
vector and tensor fields. In Sect. II we shall construct a quantum description for spacetime in a new Weylian-like manifold, here introduced. In Sect. III, we shall study the example of a Reissner-Nordstr\"om black-hole, and we shall made some remarks on the quantum geometrical structure of a Schwarzschild black-hole. Finally, in Sect. IV we shall include some remarks.

\subsection{Variation of EH action}

It is known that in the event that a manifold has a boundary
$\partial{\cal{M}}$, the action should be supplemented by a
boundary term so that the variational principle to be
well-defined\cite{Y,GH}. However, this is not the only manner to
study this problem. As was recently demonstrated\cite{RB}, there
is another way to include the flux around a hypersurface that
encloses a physical source without the inclusion of another term
in the Einstein-Hilbert (EH) action, but by making a constraint
on the first variation of the EH action. En that paper was
demonstrated that the non-zero flux of the vector metric
fluctuations through the closed 3D Gaussian-like hypersurface, is
responsible for the gauge-invariance of gravitational waves.

To see it, we consider the problem of a EH action ${\cal I}$, which describes gravitation and matter
\begin{equation}\label{act}
{\cal I} =\int_V d^4x \,\sqrt{-g} \left[ \frac{R}{2\kappa} + {\cal L}_m\right].
\end{equation}
The first term in (\ref{act}) is the Einstein-Hilbert action and $\kappa = 8 \pi G$.
Here, $g$ is the determinant of the covariant background tensor metric $g_{\mu\nu}$, $R=g^{\mu\nu} R_{\mu\nu}$ is the scalar curvature,
$R^{\alpha}_{\mu\nu\alpha}=R_{\mu\nu}$ is
the covariant Ricci tensor and ${\cal L}_m$ is an arbitrary Lagrangian density which describes matter. If we deal with an orthogonal base,
the curvature tensor will be written in terms of the connections:
$R^{\alpha}_{\,\,\,\beta\gamma\delta} = \Gamma^{\alpha}_{\,\,\,\beta\delta,\gamma} -  \Gamma^{\alpha}_{\,\,\,\beta\gamma,\delta}
+ \Gamma^{\epsilon}_{\,\,\,\beta\delta} \Gamma^{\alpha}_{\,\,\,\epsilon\gamma} - \Gamma^{\epsilon}_{\,\,\,\beta\gamma}
\Gamma^{\alpha}_{\,\,\,\epsilon\delta}$.

The first variation of the action is
\begin{equation}\label{delta}
\delta {\cal I} = \int d^4 x \sqrt{-g} \left[ \delta g^{\alpha\beta} \left( G_{\alpha\beta} + \kappa T_{\alpha\beta}\right)
+ g^{\alpha\beta} \delta R_{\alpha\beta} \right],
\end{equation}
with $g^{\alpha\beta} \delta R_{\alpha\beta} =\nabla_{\alpha}
\delta W^{\alpha}$, where  $\delta W^{\alpha}=\delta
\Gamma^{\alpha}_{\beta\gamma} g^{\beta\gamma}-
\delta\Gamma^{\epsilon}_{\beta\epsilon}
g^{\beta\alpha}=g^{\beta\gamma} \nabla^{\alpha}
\delta\Psi_{\beta\gamma}$\cite{HE}. When we deal with a manifold
${\cal M}$ which has a boundary $\partial {\cal M}$, the action
(\ref{act}) should be supplemented by a boundary term in order to
the variational principle to be well-defined. This additional term
is known as the York-Gibbons-Hawking action\cite{Y,GH}.

\subsection{Gauge transformations}

We shall propose other solution for this problem. Our
strategy will be to preserve the Einstein-Hilbert action as in
(\ref{act}) and see what are the consequences of do it. In order to make
$\delta {\cal I}=0$ in (\ref{delta}), we must consider the condition: $
G_{\alpha\beta} + \kappa T_{\alpha\beta} = \Lambda\,
g_{\alpha\beta}$, where $\Lambda$ is the cosmological constant. Additionally, we must require that
$g^{\alpha\beta} \delta R_{\alpha\beta} =\nabla_{\alpha}
\delta W^{\alpha} = \delta\Phi$, so that we obtain
the constraint $ g_{\alpha\beta}\,\delta\Phi =\Lambda\,\delta g_{\alpha\beta} $, where we have used: $g^{\alpha\beta} \delta g_{\alpha\beta}=-g_{\alpha\beta} \delta g^{\alpha\beta}$. Then, we propose the existence of a tensor
field $\delta\Psi_{\alpha\beta}$, such that $\delta
R_{\alpha\beta}\equiv \nabla_{\beta} \delta W_{\alpha}-\delta\Phi
\,g_{\alpha\beta} \equiv \Box \delta\Psi_{\alpha\beta} -\delta\Phi
\,g_{\alpha\beta} =- \kappa \,\delta S_{\alpha\beta}$\footnote{We
have introduced the tensor $S_{\alpha\beta} = T_{\alpha\beta}
-\frac{1}{2} T \, g_{\alpha\beta}$, which takes into account
matter as a source of the Ricci tensor $R_{\alpha\beta}$.}, and
hence $\delta W^{\alpha} = g^{\beta\gamma} \nabla^{\alpha}
\delta\Psi_{\beta\gamma}$, with $\nabla^{\alpha}
\delta\Psi_{\beta\gamma}=\delta\Gamma^{\alpha}_{\beta\gamma} -
\delta^{\alpha}_{\gamma} \delta\Gamma^{\epsilon}_{\beta\epsilon}$.
{\em Notice that the fields $\bar{ \delta W}_{\alpha}$ and
$\bar{\delta\Psi}_{\alpha\beta}$ are gauge-invariant under
transformations}:
\begin{equation}
\bar{\delta W}_{\alpha} = \delta W_{\alpha} - \nabla_{\alpha} \delta\Phi, \qquad
\bar{\delta\Psi}_{\alpha\beta} =\delta\Psi_{\alpha\beta} - \delta\Phi \,
g_{\alpha\beta}, \label{gauge}
\end{equation}
where $\delta\Phi$ complies $\Box \delta\Phi =0$. This means that exists a family of vector and tensor fields described by (\ref{gauge}), that are related to the Einstein tensor transformations
\begin{equation}\label{ein}
\bar{G}_{\alpha\beta} = {G}_{\alpha\beta} - \Lambda\, g_{\alpha\beta},
\end{equation}
and leave invariant the action. The transformed Einstein equations with the equation of motion for the transformed gravitational waves, hold
\begin{eqnarray}
&& \bar{G}_{\alpha\beta} = - \kappa\, {T}_{\alpha\beta}, \label{e1} \\
&& \Box \bar{\delta\Psi}_{\alpha\beta} =- \kappa \,\delta
S_{\alpha\beta}, \label{e2}
\end{eqnarray}
with $\Box \delta\Phi(x^{\alpha})=0$ and\footnote{This expression is equivalent to
$\Lambda\, \delta{g}_{\alpha\beta} = g_{\alpha\beta}\, \delta{\Phi}$.}
\begin{equation}\label{ccon}
\delta\Phi(x^{\alpha}) =  \frac{\Lambda}{4}\,g^{\alpha\beta}\,\delta g_{\alpha\beta}.
\end{equation} 
The eq. (\ref{e1}) provides us the Einstein equations with cosmological
constant included, and (\ref{e2}) describes the exact equation of
motion for gravitational waves with an arbitrary physical source $\delta
S_{\alpha\beta}$ inside a Gaussian-like hypersurface. A very
important fact is that the scalar field $\delta\Phi(x^{\alpha})$ appears
as a scalar flux of the tetra-vector with components $\delta W^{\alpha}$
through the closed hypersurface $\partial{\cal M}$. This arbitrary
hypersurface encloses the manifold by down and must be viewed as a
3D Gaussian-like hypersurface situated in any region of
space-time. This scalar flux is a gravitodynamic potential related
to the gauge-invariance of $\delta W^{\alpha}$ and the gravitational
waves $\bar{\delta\Psi}_{\alpha\beta}$. These waves appear by varying the Ricci tensor, as in the case of a flat background. 
However, in our case this variation is exact and was done in an arbitrary background. Other important fact is
that since $\delta \Phi(x^{\alpha}) = \frac{\Lambda}{4}\, g^{\alpha\beta} \delta
g_{\alpha\beta}$, the existence of the Hubble horizon is related
to the existence of the Gaussian-like hypersurface with an inner source. The variation of the
metric tensor must be done in a Weylian-like
integrable manifold\cite{RB} using an
auxiliary geometrical scalar field $\theta$, in order to the
Einstein tensor (and the Einstein equations) can be represented on
a Weyl-like manifold, in agreement with the gauge-invariant
transformations (\ref{gauge}).

In this letter we shall explore the possibility  that the
variation of the tensor metric must be done in a Weylian-like
integrable manifold (defined in the next section) using an
auxiliary geometrical scalar field $\sigma$, in order to the
Einstein tensor (and the Einstein equations) can be represented on
a Weyl manifold, in agreement with the gauge-invariant
transformations (\ref{gauge}). We shall study the relativistic
quantum dynamics of $\sigma$ by using the fact that $\Lambda$ is a
relativistic invariant. Finally, to illustrate the formalism, we shall work an example of a Reissner-Nordstr\"om (RN)
black-hole in order to obtain the dynamic equation for gravitational waves.

\section{Weylian-like representation of the Einstein tensor}

In the sense of the Riemannian geometry, the covariant derivative is null, so that $\Delta g_{\alpha\beta}=g_{\alpha\beta;\gamma} \,dx^{\gamma}=0$,
where we denote with $;$ the Riemann-covariant derivative. The Weyl geometry\cite{Weyl} is a generalization of the Riemannian geometry. In this letter
we shall consider an alternative proposal to the Weyl covariant derivative, in which the metric tensor is also nonzero: $ g_{\alpha\beta|\gamma} = \phi_{,\gamma}\,g_{\alpha\beta}$. Here, the " $|\gamma $ " denotes the new Weyl-like covariant derivative with respect to the Weyl-like\cite{Weyl} connections $\Gamma^{\alpha}_{\beta\gamma}$, given by\footnote{To simplify the notation we shall denote $\sigma_{\alpha} \equiv \sigma_{,\alpha}$}\footnote{The connections (\ref{ga}) could be generalised to other in which the torsion to be nonzero, but this issue will be studied in a future work.}
\begin{equation}\label{ga}
\Gamma^{\alpha}_{\beta\gamma} = \left\{ \begin{array}{cc}  \alpha \, \\ \beta \, \gamma  \end{array} \right\}+ g_{\beta\gamma} \sigma^{\alpha}.
\end{equation}
This connections are very similar to the Weyl ones. In both cases the non-metricity in nonzero.
The variation of the metric tensor in the sense of (\ref{ga})\footnote{In what follows we shall denote with a $\Delta$ variations on the Riemann manifold, and with a $\delta$ variations on a Weylian-like manifold.}: $\delta g_{\alpha\beta}$, will be\footnote{The reader can see using the constraint (\ref{ccon}) and (\ref{ga}) that $\delta\Phi=-{\Lambda \over 2} d\sigma$.}
\begin{equation}\label{gab}
\delta g_{\alpha\beta} = g_{\alpha\beta|\gamma} \,dx^{\gamma} = -\left[\sigma_{\beta} g_{\alpha\gamma} +\sigma_{\alpha} g_{\beta\gamma}
\right]\,dx^{\gamma},
\end{equation}
where
\footnote{We can define the operator
\begin{displaymath}
\check{x}^{\alpha}(t,\vec{x}) = \frac{1}{(2\pi)^{3/2}} \int d^3 k \, \check{e}^{\alpha} \left[ b_k \, \check{x}_k(t,\vec{x}) + b^{\dagger}_k \, \check{x}^*_k(t,\vec{x})\right],
\end{displaymath}
such that $b^{\dagger}_k$ and $b_k$ are the creation and destruction operators of space-time, such that $\left< B \left| \left[b_k,b^{\dagger}_{k'}\right]\right| B  \right> = \delta^{(3)}(\vec{k}-\vec{k'})$ and $\check{e}^{\alpha}=\epsilon^{\alpha}_{\,\,\,\,\beta\gamma\delta} \check{e}^{\beta} \check{e}^{\gamma}\check{e}^{\delta}$. }
\begin{equation}
dx^{\alpha} \left. | B \right> =  \hat{U}^{\alpha} dS \left. | B \right>= \delta\check{x}^{\alpha} (x^{\beta}) \left. | B \right> ,
\end{equation}
is the eigenvalue that results when we apply the operator $ \delta\check{x}^{\alpha} (x^{\beta}) $ on a background quantum state $ \left. | B \right> $, defined on the Riemannian manifold\footnote{In our case the background quantum state can be represented in a ordinary Fock space in contrast with LQG, where operators is qualitatively different
from the standard quantization of gauge fields.}. We shall denote with a " hat " the quantities represented on the Riemannian background manifold. The Weylian-like line element is given by
\begin{equation}
dS^2 \, \delta_{BB'}=\left( \hat{U}_{\alpha} \hat{U}^{\alpha}\right) dS^2\, \delta_{BB'} = \left< B \left|  \delta\check{x}_{\alpha} \delta\check{x}^{\alpha}\right| B'  \right>.
\end{equation}
Hence, the differential Weylian-like line element $dS$ provides the displacement of the quantum trajectories with respect to the "classical" (Riemannian) ones. When we displace
with parallelism some vector $v^{\alpha}$ on the Weylian-like manifold, we obtain
\begin{equation}
\delta v^{\alpha} = \sigma^{\alpha} g_{\beta\gamma} v^{\beta} dx^{\gamma}, \qquad \rightarrow \qquad \frac{\delta v^{\alpha}}{\delta S} =
\sigma^{\alpha} v^{\beta}\,g_{\beta\gamma}\, \hat{U}^{\gamma},
\end{equation}
where we have taken into account that the variation of $v^{\alpha}$ on the Riemannian manifold, is zero: $\Delta v^{\alpha}=0$. Hence, the norm of the vector on the Weylian-like
manifold is not conserved: $\frac{\delta v^{\alpha}}{\delta S} \frac{\delta v_{\alpha}}{\delta S}=-\left(\sigma^{\alpha} \hat{U}_{\alpha}\right) \left(v_{\gamma} \hat{U}^{\gamma} \right) \left(\sigma^{\nu} v_{\nu}\right) \neq 0$.
From the action's point of view, the scalar field $\sigma(x^{\alpha})$ drives a geometrical displacement from a Riemannian manifold to a Weylian-like one, that leaves the action invariant
\begin{equation}
{\cal I} = \int d^4 x\, \sqrt{-\hat{g}}\, \left[\frac{\hat{R}}{2\kappa} + \hat{{\cal L}}\right] = \int d^4 x\, \left[\sqrt{-\hat{g}} e^{-2\sigma}\right]\,
\left\{\left[\frac{\hat{R}}{2\kappa} + \hat{{\cal L}}\right]\,e^{2\sigma}\right\}.
\end{equation}
If we require that $\delta {\cal I} =0$, we obtain
\begin{equation}
-\frac{\delta V}{V} = \frac{\delta \left[\frac{\hat{R}}{2\kappa} + \hat{{\cal L}}\right]}{\left[\frac{\hat{R}}{2\kappa} + \hat{{\cal L}}\right]}
= 2 \,\delta\sigma,
\end{equation}
where $\delta\sigma = \sigma_{\mu} dx^{\mu}$ is an exact differential and $V=\sqrt{-\hat{ g}}$ is the volume of the Riemannian manifold. Of course, all the variations are in the Weylian-like geometrical representation, and assure us gauge invariance because $\delta {\cal I} =0$.

\subsection{Gauge-invariant relativistic dynamics on a Weylian-like manifold}

The Ricci tensor in the Weylian-like and Riemann representations can be related by
\begin{equation}
\bar{R}_{\alpha\beta} =\hat{R}_{\alpha\beta} +  \sigma_{\alpha ; \beta} + \sigma_{\alpha} \sigma_{\beta} - g_{\alpha\beta}
\left[ \left(\sigma^{\mu}\right)_{;\mu} + \sigma_{\mu} \sigma^{\mu} \right],
\end{equation}
so that both representations of the scalar curvature, are related by
\begin{equation}
\bar{R} = \hat{R} - 3\left[\left( \sigma^{\mu}\right)_{;\mu} + \sigma_{\mu} \sigma^{\mu} \right].
\end{equation}
The Einstein tensor can be written as
\begin{equation}
\bar{G}_{\alpha\beta} = \hat{G}_{\mu\nu} + \sigma_{\alpha ; \beta} + \sigma_{\alpha} \sigma_{\beta} + \frac{1}{2} \,g_{\alpha\beta}
\left[ \left(\sigma^{\mu}\right)_{;\mu} + \sigma_{\mu} \sigma^{\mu} \right],
\end{equation}
where we have made use of the fact that the connections are symmetric.

\subsection{Weylian action and quantum algebra}

Now we consider the expression (\ref{ein}), from which we obtain
that the invariant $\Lambda$. From the point of view of the Riemann manifold $\Lambda$ is a constant, but from the point of view of the Weylian-like manifold: $\Lambda\equiv \Lambda(\sigma, \sigma_{\alpha})$ can be considered a functional, given by
\begin{equation}\label{aa}
\Lambda(\sigma, \sigma_{\alpha}) = -\frac{3}{4} \left[ \sigma_{\alpha} \sigma^{\alpha} + \hat{\Box} \sigma\right].
\end{equation}
Therefore, we can define a geometrical quantum action on the Weylian-like manifold with (\ref{aa})
\begin{equation}
{\cal W} = \int d^4 x \, \sqrt{-g} \,\, \Lambda(\sigma, \sigma_{\alpha}),
\end{equation}
such that the dynamics of the geometrical field is given by the Euler-Lagrange equations, after imposing $\delta
W=0$:
\begin{equation}
\frac{\delta \Lambda}{\delta \sigma} - \hat{\nabla}_{\alpha} \left( \frac{\delta \Lambda}{\delta \sigma_{\alpha}}\right) =0,
\end{equation}
where the variations are defined on the Weylian-like manifold. This means that $\delta\Lambda\neq 0$, but $\Delta\Lambda=0$. Furthermore, $ \Pi^{\alpha}=\frac{\delta \Lambda}{\delta \sigma_{\alpha}}=-{3\over 4} \sigma^{\alpha}$ is the geometrical momentum and the
dynamics of $\sigma$ describes a free scalar field
\begin{equation}\label{si}
\hat{\Box} \sigma =0,
\end{equation}
so that the momentum components $\Pi^{\alpha}$ comply with
the equation
\begin{equation}
\hat{\nabla}_{\alpha} \Pi^{\alpha} =0.
\end{equation}
If we define the scalar invariant
$\Pi^2=\Pi_{\alpha}\Pi^{\alpha}$, we obtain that
\begin{equation}
\left[\sigma,\Pi^{2}\right] = \frac{9}{16}\left\{ \sigma_{\alpha} \left[\sigma,\sigma^{\alpha} \right]
 + \left[\sigma,\sigma_{\alpha} \right] \sigma^{\alpha} \right\}=0,
\end{equation}
where we have used that $\sigma_{\alpha} U^{\alpha} = U_{\alpha} \sigma^{\alpha}$, and
\begin{equation}\label{con}
\left[\sigma(x),\sigma^{\alpha}(y) \right] =- i \Theta^{\alpha}\, \delta^{(4)} (x-y), \qquad \left[\sigma(x),\sigma_{\alpha}(y) \right] =
i \Theta_{\alpha}\, \delta^{(4)} (x-y),
\end{equation}
with $\Theta^{\alpha} = \hbar\, \hat{U}^{\alpha}$. Therefore we can define
the relativistic invariant $\Theta^2 = \Theta_{\alpha}
\Theta^{\alpha} = \hbar^2 \hat{U}_{\alpha}\, \hat{U}^{\alpha}$,
where $\hat{U}^{\alpha}$ are the components of the Riemannian
velocities. Additionally, it is possible to define the Hamiltonian operator
\begin{equation}
{\cal H} = \left(\frac{\delta \Lambda}{\delta \sigma_{\alpha}}\right) \sigma_{\alpha} - \Lambda(\sigma,\sigma_{\alpha}),
\end{equation}
such that the eigenvalues of "quantum energy" becomes from ${\cal H}\left|B\right> = E\left|B\right>$. Can be demonstrated that
$\delta{\cal H}=0$, so that the quantum energy $E$ is a Weylian-like invariant.

\subsection{Gravitational waves with Weylian-like variations}

Now we consider the Weylian-like variation of the Ricci tensor: $\delta R_{\alpha\beta}$. This is given by
\begin{equation}\label{gw}
\hat\Box \delta\Psi_{\alpha\beta} = -\kappa
\,\delta\left(\hat{L}_{\alpha\beta}\right),
\end{equation}
where $\hat\Box \delta\Psi_{\beta\gamma} \equiv \hat\nabla_{\alpha}
\hat\nabla^{\alpha} \delta\Psi_{\beta\gamma} = \hat\nabla_{\alpha} \left(
\delta\Gamma^{\alpha}_{\beta\gamma} - \delta^{\alpha}_{\gamma}
\delta \Gamma^{\epsilon}_{\beta\epsilon}\right)$,
$\delta\hat{L}_{\alpha\beta}=\delta\left(\frac{\delta \hat{\cal
L}}{\delta g^{\alpha\beta}}\right)$, such that $\hat{{\cal L}} =
 g^{\alpha\beta} \hat{L}_{\alpha\beta}$, and we have
considered that $\delta I=0$\cite{RB}. On Riemannian  hypersurfaces all the field solutions are
background solutions, so that we can consider that $\Delta
\left(\hat{L}_{\alpha\beta}\right)=0$. In this case we obtain
\begin{equation}\label{gw1}
\hat\Box \delta\Psi_{\alpha\beta} = 2\kappa \, \sigma^{\theta}
\hat{L}_{\theta\beta}\, dx_{\alpha},
\end{equation}
where $dx_{\alpha} = \hat{U}_{\alpha} dS$, such that if we rename
$\chi_{\alpha\beta} ={\delta\Psi_{\alpha\beta}\over \delta S}$, we
finally obtain
\begin{equation}\label{gw}
\hat\Box \chi_{\alpha\beta} = 2\kappa \, \sigma^{\theta}
\hat{L}_{\theta\beta}\,\hat{U}_{\alpha},
\end{equation}
which describes the Riemannian dynamics of gravitational waves on a
Weylian-like hypersurface\footnote{It is possible to show that exists a Weylian-like dynamics of $\sigma$ given by the Euler-Lagrange equation $\frac{\delta \Lambda}{\delta \sigma} - \nabla_{\alpha}\left( \frac{\delta \Lambda}{\delta \sigma_{\alpha}}\right) =0$, such that $\nabla_{\alpha}$ denotes the Weylian-like covariant derivative. Can be demonstrated that
the solutions of $\sigma$ in the Weylian-like gauge can be expressed in terms of a Fourier expansion of travelling waves that moves with light velocity. This means that the solution
of $\chi_{\alpha\beta}(x^{\alpha})$ must be written as a superposition of travelling waves in the Weylian-like manifold.}. Notice that the source depends on how the relativistic observer is moving on the
Riemannian hypersurface, which is determined by $\hat{U}_{\alpha}$. A very important fact is that the source in
the equation (\ref{gw}) depends on the field $\sigma^{\theta}$. The existence of this field in the source is intrinsically related to the existence of the scalar flux $\Phi$, and the cosmological constant $\Lambda$. This can be
seen by noting that
\begin{equation}
\frac{\delta \sigma}{\delta S} = \sigma_{\gamma} \hat{U}^{\gamma} = - \frac{2}{\Lambda} \frac{\delta \Phi}{\delta S},
\end{equation}
where we have used the constraint $\delta g_{\alpha\beta} \Lambda =
\delta\Phi\, g_{\alpha\beta}$ and (\ref{gab}). In the next section we shall consider
the example of an RN black-hole, to obtain gravitational waves.

\section{An example: RN black hole}

We consider a RN black-hole, with mass $\bar{M}=2 G M$, and squared
electric charge $Q^2$, such that the line element is given by
\begin{equation}\label{bh}
dS^2 = f(r) dt^2 - \frac{1}{f(r)} dr^2 - r^2\, d\Omega^2,
\end{equation}
where $d\Omega^2 = \sin^2\theta\,d\phi^2 + d\theta^2$ is the
square differential of solid angle and $f(r) = 1-{\bar{M}\over r}
+ {Q^2\over r^2}$, such that $\bar{M}=2 G M$ ($M$ is the mass of
the charged black-hole), $\bar{M}^2 \geq (2Q)^2$ and $Q={q\over
4\pi \epsilon_0}$. The horizon radius is given by
\begin{equation}
r_h=\frac{\bar{M}}{2} \left[1-\left(1-\left(\frac{4Q}{\bar{M}}\right)^2\right)^{1/2}\right].
\end{equation}

In this case the Lagrangian density is given by\cite{rb}
\begin{equation}
\hat{\cal L} = \frac{1}{2 \, f(r)} \left[\vec{E}.\vec{E} -
\vec{\nabla}\phi(r).\vec{\nabla}\phi(r)\right]=\frac{1}{2r^2\,(r^2-r \bar{M} + Q^2)} \left[ Q^2 -\frac{\bar{M}^2}{4}\right],
\end{equation}
where $|\vec{E}|^2={Q^2\over r^4}$ is the electric field due the the
charge $Q$ and $\phi(r)=-{\bar{M}\over 2 r}$ the gravitational
potential due to the mass $\bar{M}$ of the black hole described by
the line element (\ref{bh}). Therefore, the tensor density
$\hat{L}_{\alpha\beta}$ is given by
\begin{equation}
\hat{L}_{\alpha\beta} = \frac{1}{4\,f(r)}
\,g_{\alpha\beta}\,\left[\vec{E}.\vec{E} -
\vec{\nabla}\phi(r).\vec{\nabla}\phi(r)\right],
\end{equation}
so that the equation for gravitational waves holds
\begin{equation}
\hat{\Box} \chi_{\alpha\beta} = \frac{\kappa}{2\,f(r)} \sigma^{\epsilon}
\,g_{\epsilon\beta} \left[\vec{E}.\vec{E} -
\vec{\nabla}\phi(r).\vec{\nabla}\phi(r)\right]\, \hat{U}_{\alpha}.
\end{equation}
By multiplying with $g^{\alpha\beta}$, we obtain
\begin{equation}
\hat{\Box}\chi = \frac{\kappa}{2\left[f(r)\right]^{3/2}} \sigma_0 \left[|\vec{E}|^2 -
|\vec{\nabla}\phi |^2 \right] = 4 \frac{d\Phi}{dS},
\end{equation}
where
\begin{equation}
\left[|\vec{E}|^2 - |\vec{\nabla}\phi |^2 \right]=
\frac{1}{r^4} \left[ Q^2 -\frac{\bar{M}^2}{4} \right].
\end{equation}
In order to solve the equation (\ref{si}), we can make separation
of variables: $\sigma_{klm} \sim T_k(t)\,Y_{lm}
(\theta,\phi)\,R_{kl}(r)$. After making $\omega=k\, c$ (we use
$c=1$)\footnote{Here, $\omega$ and $k$ are respectively the
frequency and the wavenumber in the coordinated system.}, we
obtain the solutions
\begin{eqnarray}
T_{k}(t) &=& e^{\pm i k\,c \, t}, \\
Y_{lm} (\theta,\phi) &=& \sqrt{\frac{(2l+1)(l-m)\!}{4\pi (l+m)\!}} \, e^{i m \phi} \, P_{lm} (\cos{\theta}), \\
R_{kl}(r) &=& \frac{e^{i k r}}{r} \left\{ C_1\,\left(2 r -\bar{M} -\sqrt{\bar{M}^2-4Q^2}\right)^A \, \left( \bar{M} - 2r -\sqrt{\bar{M}^2
-4Q^2}\right)^B \right.\nonumber \\
&\times & \, {\cal H_C}\left[ 2i k \sqrt{\bar{M}^2-4Q^2}, \alpha_1,
\beta_1, 2 \bar{M} k^2 \sqrt{\bar{M}^2-4Q^2}, \gamma_1, \frac{2r-\bar{M} +
\sqrt{\bar{M}^2-4Q^2}}{2 \sqrt{\bar{M}^2-4Q^2}}
\right] \nonumber \\
& + &  C_2 \, \left( r - \frac{\bar{M}}{2} - \frac{\sqrt{\bar{M}^2-4Q^2}}{2} \right)^B \, \left(2 r - \bar{M}
+ \sqrt{\bar{M}^2-4Q^2}\right)^C \nonumber \\
 & \times & \left.{\cal H_C}\left[
2i k \sqrt{\bar{M}^2-4Q^2}, -\alpha_1, \beta_1, 2 \bar{M} k^2 \sqrt{\bar{M}^2-4Q^2},
\gamma_1, \frac{2r-\bar{M} + \sqrt{\bar{M}^2-4Q^2}}{2 \sqrt{\bar{M}^2-4Q^2}}
\right]\right\} \nonumber \\
\end{eqnarray}
where ${\cal H_C}$ are the confluent Heun functions, ($C_1, C_2$) are constants and
\begin{small}
\begin{eqnarray}
A & = & \frac{\sqrt{\bar{M}^2-4Q^2}+\left[4Q^2 k^2 \bar{M}^2 \left( 2 -\frac{Q^2}{\bar{M}^2} - \sqrt{1-4\frac{Q^2}{\bar{M}^2}}\right)-2
\bar{M}^4 k^2+\sqrt{\bar{M}^2-4Q^2}
\bar{M}^3 k^2 +\bar{M}^2-4Q^2\right]^{1/2}}{2\sqrt{\bar{M}^2-4Q^2}}, \\
B & = & \frac{4Q^2-\bar{M}^2+\sqrt{\bar{M}^2-4Q^2}\left[4Q^2 k^2 \bar{M}^2 \left( 2 -\frac{Q^2}{\bar{M}^2} - \sqrt{1-4\frac{Q^2}{\bar{M}^2}}\right)-
2 \bar{M}^4 k^2-\sqrt{\bar{M}^2-4Q^2} \bar{M}^3 k^2 +\bar{M}^2-4Q^2\right]^{1/2}}{2(4Q^2-\bar{M}^2)}, \nonumber \\
&& \\
C & = & \frac{4Q^2-\bar{M}^2+\sqrt{\bar{M}^2-4Q^2}\left[4Q^2 k^2 \bar{M}^2 \left( 2 -\frac{Q^2}{\bar{M}^2} + \sqrt{1-4\frac{Q^2}{\bar{M}^2}}\right)-
2 \bar{M}^4 k^2-\sqrt{\bar{M}^2-4Q^2} \bar{M}^3 k^2 +\bar{M}^2-4Q^2\right]^{1/2}}{2(4Q^2-\bar{M}^2)}, \nonumber \\
&& \\
\alpha_1 & = &  \frac{\left[4Q^2 k^2 \bar{M}^2 \left( 2 -\frac{Q^2}{\bar{M}^2} + \sqrt{1-4\frac{Q^2}{\bar{M}^2}}\right)-2 \bar{M}^4 k^2-\sqrt{\bar{M}^2
-4Q^2}
\bar{M}^3 k^2 +\bar{M}^2-4Q^2\right]^{1/2}}{\sqrt{\bar{M}^2-4Q^2}},\\
\beta_1 & = & -\frac{\left[4Q^2 k^2 \bar{M}^2 \left( 2 -\frac{Q^2}{\bar{M}^2} + \sqrt{1-4\frac{Q^2}{\bar{M}^2}}\right)-2 \bar{M}^4 k^2\left(1-\sqrt{1
-4\frac{Q^2}{\bar{M}^2}}\right) +\bar{M}^2-4Q^2\right]^{1/2}}{\sqrt{\bar{M}^2-4Q^2}}, \\
\gamma_1 & = & -\frac{\left[\bar{M}^2-4Q^2-4Q^2 k^2 \bar{M}^2 \left( 3 -3\frac{Q^2}{\bar{M}^2} - 2\sqrt{1-4\frac{Q^2}{\bar{M}^2}}\right)+2 \bar{M}^4
k^2\left(1-\sqrt{1-4\frac{Q^2}{\bar{M}^2}}\right) +8Q^2 l(l+1)-2\bar{M}^2 l(l+1)\right]}{2(\bar{M}^2-4Q^2)}. \nonumber \\
\end{eqnarray}
\end{small}
Therefore, the radial function ${R}_{kl}(r)$ can be written as ${R}_{kl}(r)=\frac{e^{i\,k\,r}}{r}\,\bar{R}_{kl}(r)$.

We can make the Fourier expansion of the geometric scalar field, $\sigma$, in spherical coordinates
\begin{equation}
\sigma(\vec r,t) =\int^{\infty}_{0} dk\,\sum_{lm} \left[A_{klm}
\sigma_{klm}(\vec r,t) +A^{\dagger}_{klm} \bar\sigma^*_{klm}(\vec
r,t)\right],
\end{equation}
where $\sum_{lm}\equiv \sum_{l=0}^{\infty}\, \sum_{m=-l}^{l}$, $k$, $l$ and $m$ are respectively the wave-numbers related to the coordinates $r$, $\theta$ and $\phi$, and
\begin{equation}
\sigma_{klm}(\vec r,t)= k^2\,R_{kl}\left(r\right)\,
Y_{lm}(\theta,\phi) \,e^{i k\,c \,t}.
\end{equation}
Here, $Y_{lm}(\theta,\phi)$ are the spherical harmonics and $P_{lm} (\cos{\theta})$ are the Legendre polynomial. Furthermore, the annihilation and
creation operators obey the algebra
\begin{equation}
\left[A_{klm}, A^{\dagger}_{k'l'm'}\right] =
\delta^{(3)}(\vec{k}-\vec{k'}) \delta_{ll'} \delta_{m m'}, \qquad
\left[A_{klm}, A_{k'l'm'}\right]=\left[A^{\dagger}_{klm},
A^{\dagger}_{k'l'm'}\right]=0.
\end{equation}
In this example we shall consider that $U^{\alpha}\equiv
\left({1\over \sqrt{f(r)}},0,0,0\right)$, so that from eq. (\ref{con}),
we obtain that at equal times
\begin{equation}\label{coo}
\left[\sigma(t,\vec{r}),\partial^0{\sigma}(t,\vec{r}') \right] = -i
\hbar\,  \hat{U}^{0} \,\delta^{(3)} (\vec{r}-\vec{r}'),
\end{equation}
so that we obtain that $A_{klm}={1\over
(2\,k)^{3/2}}\,\sqrt{\frac{(2l+1)(l-m)\!}{4\pi (l+m)\!}}$, and
hence $\partial^0 \sigma_{klm}={\dot\sigma_{klm}\over c f(r)}=
{i\,k\over f(r)} \,\sigma_{klm}(\vec{r},t)$. Hence, in order to the commutator (\ref{coo}) to be fulfilled, we must require
that
\begin{equation}\label{cn}
\bar{R}_{kl}(r) \bar{R}^*_{kl}(r') + \bar{R}_{kl}(r') \bar{R}^*_{kl}(r)=c\, \hbar.
\end{equation}
The expression (\ref{cn}) give us the normalization condition for the modes of the Fourier expansion in any charged black-hole. The case where
the charge $Q$ is null, describes a Schwarzschild black-hole as a particular solution the case here studied.

\subsection{Schwarzschild black-hole's change of entropy}

A particular case of interest is that of $Q=0$. Making use of the fact that $\delta {\cal W}=0$, we can obtain the relationship between the scalar flux and the change of area $\delta {\cal A}$ of the black-hole
\begin{equation}
\frac{\delta\Phi}{\delta S} = \frac{3}{4} {\rm ln}\left[ \frac{\delta{\cal A}}{\bar{{\cal A}}} +1\right],
\end{equation}
where ${\delta{\cal A}\over {\bar{\cal A}}}= 2 \frac{r_s}{\bar{r}^2_s } \delta r_s$, $ \frac{{\cal A}}{{\cal A}}=\left(\frac{r_{s}}{\bar{r}_{s}}\right)^2 = e^{\frac{2}{9} \Lambda S^2} \geq 0$, and the square of the radius between the final $r_s$ and the initial Schwarzschild radius $\bar{r}_s$ is related with the change of area and the scalar flux $\Phi$, and the change of the mass's black hole: $\delta {\cal A}= \frac{4}{T} \delta{M}$ by the Bekenstein-Hawking's law\cite{bh}. Moreover, using the fact that $\delta{\cal W}=0$, and $d\sigma = -{4\over \Lambda} d\Phi$, we obtain that the change of volume of the Weylian-like manifold (the black-hole's increase of volume), with respect to the Riemannian one $\sqrt{-\hat{g}}$, is
\begin{equation}\label{vol}
V = \hat{V}\, e^{\frac{4}{\Lambda} \int d\Phi} = \hat{V}\, e^{\frac{\Lambda S^2}{3}},
\end{equation}
where $\Lambda S^2 >0$. This means that $V \geq \sqrt{-\hat{g}}$, for $S^2\geq 0$, $\Lambda >0$ and $\sigma <0$. Therefore, we require a metric's signature  $(-,+,+,+)$ in order for the
cosmological constant to be positive and a metric's signature $(+,-,-,-)$ in order to have $\Lambda \leq 0$.

\section{Final comments}

We have introduced Relativistic Quantum Geometry, which is a different approach to study the
relativistic quantum structure of spacetime. In some way our approach is in the spirit of the Isham-Butterfield's one\cite{IB}, but we have incorporated the relativistic invariants in the quantum aspects of the theory. 
The dynamics of quantum spacetime is described by $\sigma$.
This is a scalar field that drives a geometrical displacement from a Riemannian manifold to a Weylian-like one, that leaves the action invariant.
Furthermore, $\sigma$ always can be quantized
because it is a free scalar field. Other important fact is that
the quantum algebra of $\left[\sigma(x), \sigma^{\mu}(y)\right] =-i
\, \Theta^{\mu} \delta^{(4)}(x-y)$ is relativistic in nature,
because $\Theta^{\mu} \Theta_{\mu}=\hbar^2 \hat{U}^{\mu}
\hat{U}_{\mu}$ is a relativistic invariant. Due to this fact the commutators depend of how is moving the relativistic
observer. In the example here studied, the observer is static with respect to the RN black-hole, so that
$\hat{U}_{0}=1/\sqrt{f(r)}$, and therefore one obtains $\left[\sigma(x),\sigma^{0}(y) \right] =- i \frac{\hbar}{\sqrt{f(r)}}\, \delta^{(4)} (x-y)$.
However, if one were studied some example where the observer is moving with respect to the source, the commutation relationship have been
different. Furthermore, we have obtained the expression (\ref{vol}) that describes how changes the volume of the quantum manifold due to the scalar flux $\Phi$. This result can be applied to cosmology to obtain a justification for the decrease of $\Lambda$ with the expansion of the universe. It is important to notice that $\Lambda$ is a Riemannian constant ($\Delta \Lambda=0$), but variate in the Weylian-like manifold ($\delta \Lambda \neq 0$).

 \section*{Acknowledgements}

\noindent The authors acknowledge M. A. R. Arcod\'{\i}a for his interesting comments, and also acknowledge
UNMdP and  CONICET (Argentina) for financial support.

\end{document}